\documentclass[aps,floatfix,prd]{revtex4}

\usepackage{graphicx}
\usepackage{xspace}

\def\kev {\,\mbox{keV}}
\def\mev {\,\mbox{MeV}}
\def\gev {\,\mbox{GeV}}
\def\pb {\,\mbox{pb}}
\def\invfb {\,\ensuremath{\mbox{fb}^{-1}}}

\begin{document}
\newcommand{\BABARPubYear}    {05}
\newcommand{\BABARPubNumber}  {071}
\newcommand{\SLACPubNumber} {11528}

\begin{flushleft}
hep-ex/xxxxxxxx\\
BABAR-PROC-\BABARPubYear/\BABARPubNumber \\
SLAC-PUB-\SLACPubNumber
\end{flushleft}

\title{ 
Inclusive Hadronic Results at BaBar :\\
ISR and Pentaquark Searches \footnote{Work supported by Department of Energy contract DE-AC02-76SF00515.}
}
\author{ Nicolas Berger, on behalf of the BaBar collaboration}

\affiliation{SLAC, Menlo Park, California, U.S.A.}

\begin{abstract}
We present recent measurements of hadronic cross-sections from the BaBar experiment and report 
preliminary results on searches for pentaquark states.
\end{abstract}

\maketitle

\section{Inclusive hadronic cross-section measurements using Initial State Radiation}
\subsection{Physics Motivation}

The total cross-section $\sigma(e^+ e^- \to \mbox{hadrons})$ for the production of hadrons
in $e^+e^-$ annihilation is a crucial ingredient for the calculation of hadronic 
corrections for the running of the QED coupling constant $\Delta\alpha_{QED}^{Had}$
and for the muon anomalous magnetic moment $a_{\mu}^{Had}$.
The hadronic contribution to the running of $\alpha_{QED}$, $\Delta\alpha_{QED}^{Had}$
is an input into the global standard model fits\cite{smfits} which can provide an
indirect measurement of the Higgs boson mass. 
In both cases, the hadronic contributions can be expressed as integrals of the ratio
$R(s) = \sigma(e^+ e^- \to \mbox{hadrons})/\sigma_0(e^+ e^- \to \mu^+ \mu^-)$, where $\sigma_0$
denotes the Born cross-section. We have

\begin{eqnarray}
\Delta\alpha_{QED}^{Had} = -\frac{\alpha}{3\pi} \int_{4 m_{\pi}^2}^{\infty}{\frac{R(s)}{s}
\left( \frac{m_Z^2}{s - m_Z^2} \right)} \\
a_{\mu}^{Had} = \left(\frac{\alpha m_{\mu}}{3\pi}\right)^2 \int_{4 m_{\pi}^2}^{\infty}{\frac{R(s)}{s}
\left( \frac{K(s)}{s} \right)},
\end{eqnarray}

where $K(s)$ is sharply peaked at $s = 0$. In the case of 
$\Delta\alpha_{QED}^{Had}$, the weight factor is almost independent of $s$ for small values of $s$,
so that the entire spectrum of $R(s)$ contributes to the integral.
In the case of $a_{\mu}$ the integral is dominated by the low $s$ region.

The error on $\Delta\alpha_{QED}^{Had}$ is dominated by the region
$1 \gev < \sqrt{s} < 7 \gev$. Below $1 \gev$, CMD-2 and KLOE
have measured\cite{kloecmd2} $\sigma(e^+ e^- \to \pi^+ \pi^-)$ to 
$<1\%$ accuracy. BES\cite{bes} has measured $R(s)$ in the 
range $2 \gev < \sqrt{s} < 5 \gev$ at $6\%$ accuracy, but there are no recent 
measurements in the region $1 \gev < \sqrt{s} < 2 \gev$, leading
to large uncertainties.

\subsection{Initial-state Radiation at $\Upsilon(4S)$ Energies}

The BaBar experiment operates at the PEP-II asymmetric $e^+e^-$ collider.
While PEP-II is a fixed-energy machine, initial-state radiation (ISR),
can be used to vary of the center-of-mass energy of hadron production.
The full spectrum of $s'$, the reduced center-of-mass energy, is accessible. The range 
$0 < s' < 7 \gev$ can be reached for ISR photon energies of 
$3-5.3 \gev$ in the center-of-mass system. The photon
can be detected by the BaBar electro-magnetic calorimeter (EMC) to provide a clear signature
for the event. In particular, the presence of a hard photon can separate $e^+e^-$ annihilation
events from beam-gas processes which constitutes an important source of background for 
energy-scan experiments. The hadronic system is also collimated by its recoil against 
a hard photon and the spectrum of the observed particles is also hardened, improving detection 
efficiency and reducing the dependence on the hadronization model. Requiring the ISR photon
in the sensitive part of the detector further improves the fiducial containment of the hadronic 
system. Final-state radiation (FSR) effects are expected to be small and kinematically well-separated
from ISR.
\begin{figure}
\begin{center}
\includegraphics[height=4.5cm]{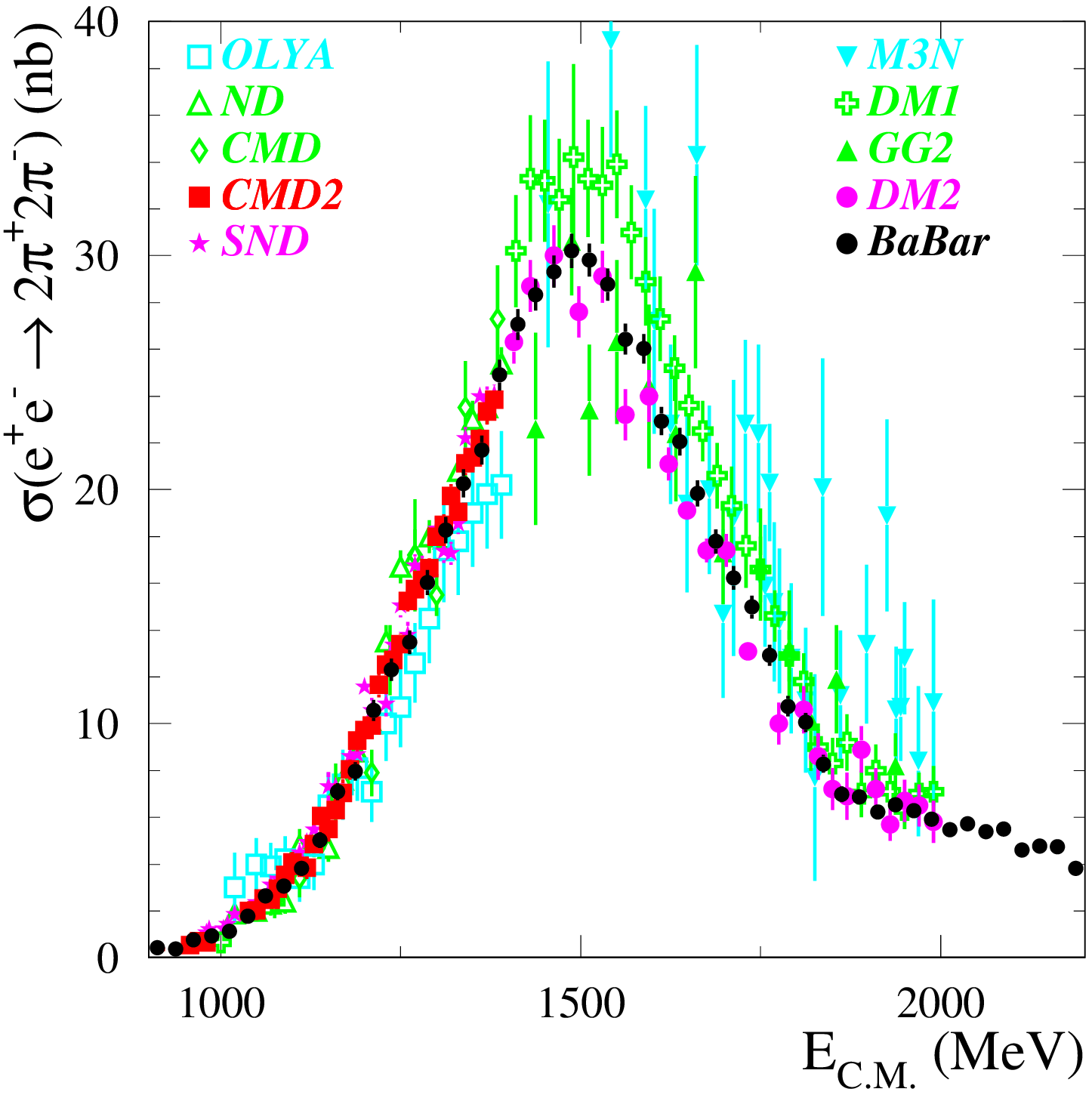}
\includegraphics[height=4.5cm]{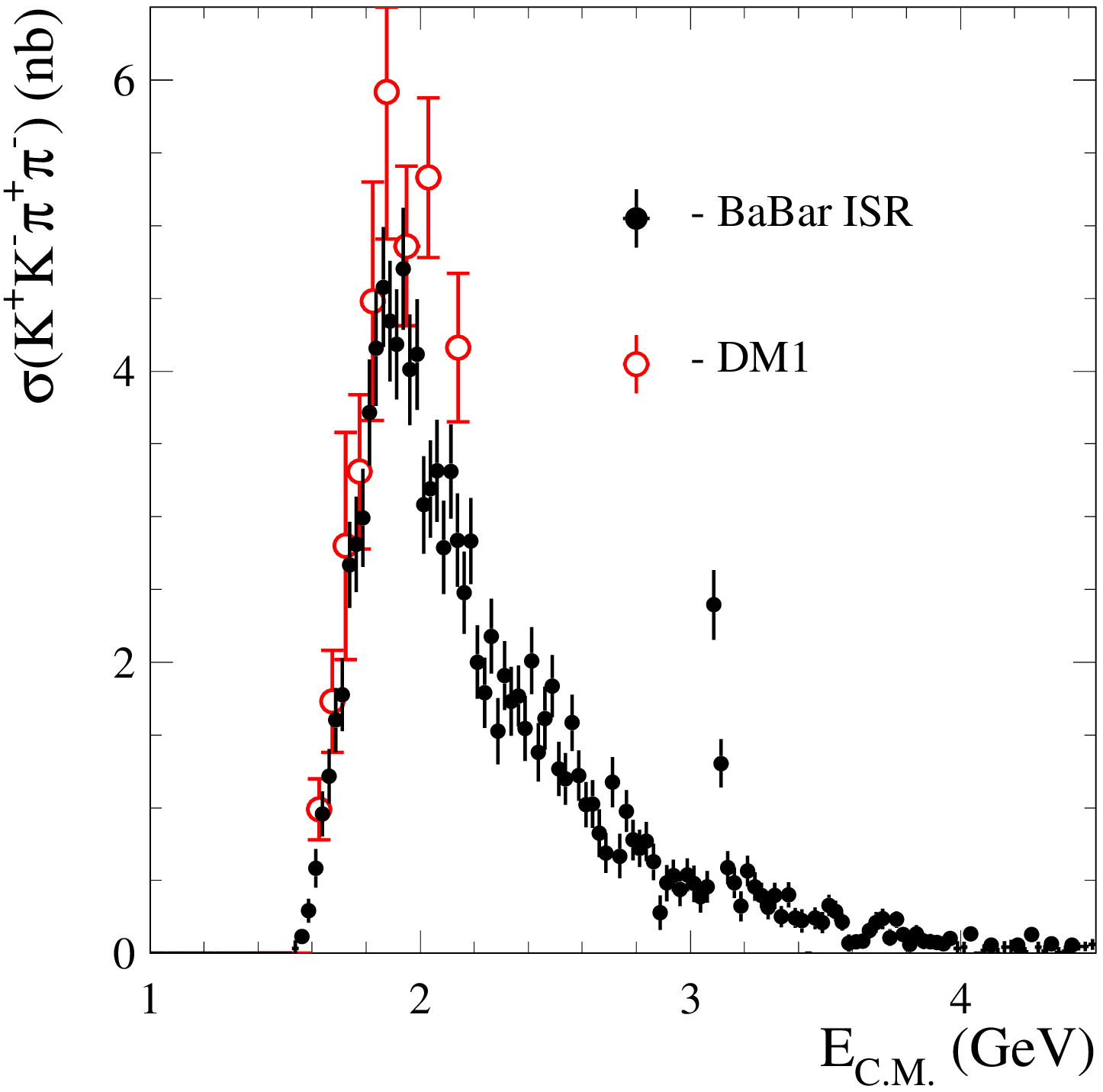}
\includegraphics[height=4.5cm]{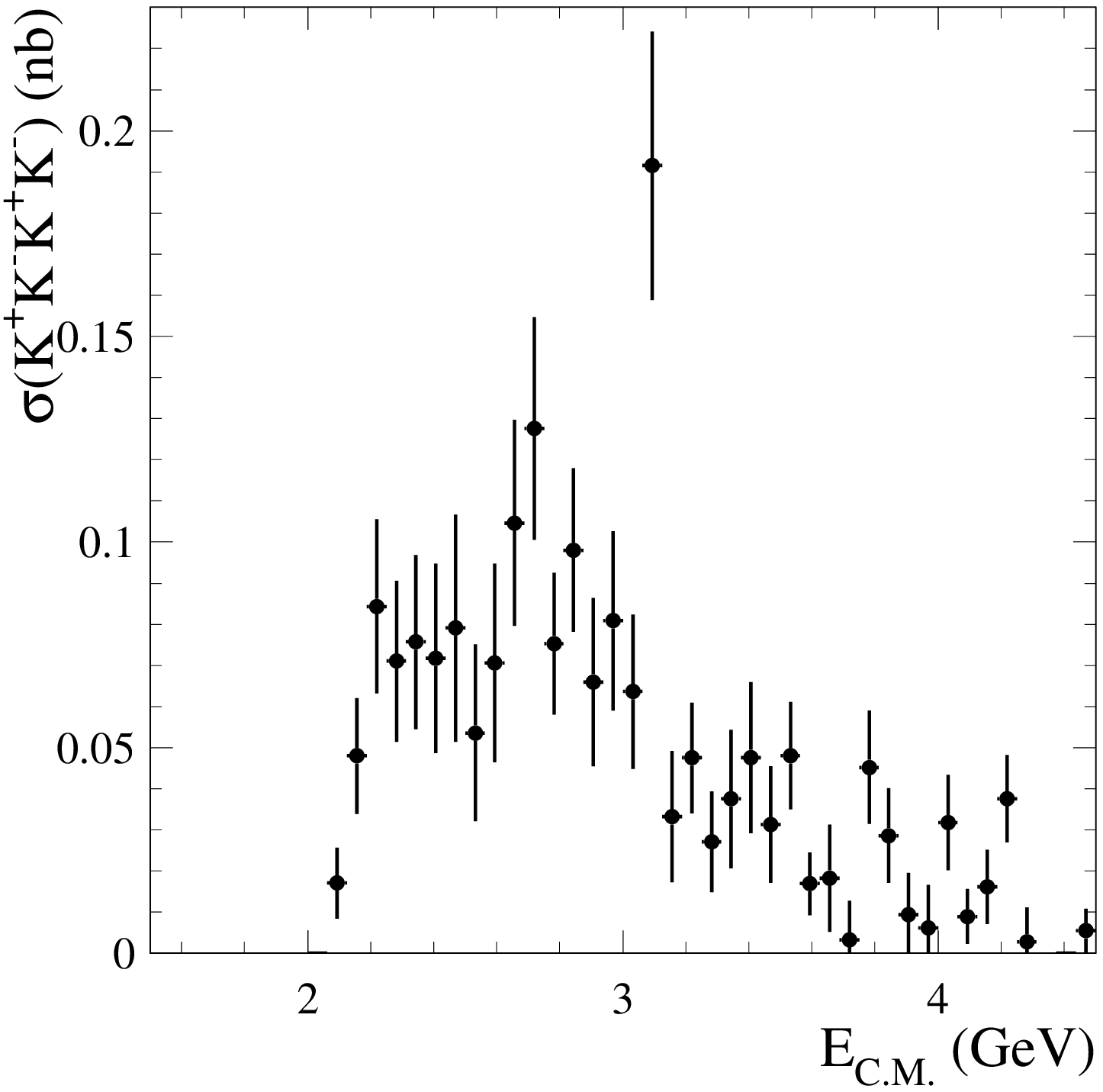}
\caption{\it Cross sections for $e^+e^- \to \pi^+ \pi^- \pi^+ \pi^-$ (top), $K^+ K^- \pi^+ \pi^-$
(center) and $K^+ K^- K^+ K^-$ (bottom) using $89 \invfb$ of data.}\label{xfourh}
\end{center}
%\vspace{-0.7cm}
\end{figure}

The cross-section for hadronic ISR events was evaluated using the Monte-Carlo generators.
The total cross-section for $s' < 8 \gev$ in the fiducial region $15.3 < \theta_{\gamma} < 137.3^o$
is calculated to be $90 \pb$, corresponding to 18 million events in the current BaBar dataset of
$200 \invfb$. Of these we expect 5.7 million events for $2 < s' < 5 \gev$, to be compared with 
approximately 250,000 events used for the latest BES measurement in this energy range.

The main challenge of the method is the determination of the reduced center-of-mass energy $\sqrt{s'}$.
This is addressed differently in the various analyses.

\subsection{$e^+e^- \to h^+ h^- h^+ h^- \gamma$}

BaBar performed a common analysis of the processes
$e^+e^- \to \pi^+ \pi^- \pi^+ \pi^- \gamma$, $e^+e^- \to K^+ K^- \pi^+ \pi^- \gamma$
and $e^+e^- \to K^+ K^- K^+ K^- \gamma$. Events with at least 4 tracks and a neutral cluster
are subjected to 1C kinematic fits with the constraint $m_{\gamma} = 0$. 
A kaon identification procedure is performed on the tracks, using ionization measurements
in the tracking detectors and information from the \u{C}erenkov detector. The cross-sections are
normalized using the process $e^+e^- \to \mu^+ \mu^- \gamma$. Results are shown in 
fig.~\ref{xfourh}.

The $4\pi$ and $2K2\pi$ results agree with existing results, but are 
considerably more precise and cover a larger energy range. The $4K$ result is the first
measurement of this quantity. In all cases, the leading uncertainties are systematic,
dominated by uncertainties on the luminosity, tracking efficiency and acceptance losses.
The $J/\psi$ resonance is clearly visible in all 3 cases, leading to branching fraction results 
of ${\cal B}(J/\psi \to \pi^+ \pi^- \pi^+ \pi^-) = (3.70 \pm 0.27 \pm 0.36) \times 10^{-3}$,
${\cal B}(J/\psi \to K^+ K^- \pi^+ \pi^-) = (6.25 \pm 0.50 \pm 0.62) \times 10^{-3}$,
${\cal B}(J/\psi \to K^+ K^- K^+ K^-) = (6.9 \pm 1.2 \pm 1.1) \times 10^{-3}$, assuming the PDG value
for $\Gamma(J/\psi \to e^+e^-)$. These results agree with the PDG values but are significantly more precise. 
The $4\pi$ mode also provides a measurement of ${\cal B}(\psi(2S) \to J/\psi(\mu^+\mu^-) \pi^+ \pi^-$ 
through the mis-identification of the muons as pions. Assuming the PDG values for $\Gamma(J/\psi \to \mu^+\mu^-)$,
$\Gamma(\psi(2S) \to e^+e^-)$ and ${\cal B}(J/\psi \to \mu^+ \mu^-)$, we get 
${\cal B}(\psi(2S) \to J/\psi \pi^+ \pi^-) = 36.1 \pm 1.5 \pm 3.7\%$.

\subsection{$e^+e^- \to J/\psi(\mu^+\mu^-) \gamma$}

The analysis of $e^+e^- \to J/\psi(\mu^+\mu^-) \gamma$ is done in similar fashion to that of the
preceding section. We require energy and momentum conservation and perform a
1C kinematic fit with $m_{\gamma} = 0$. To reject ISR background, both tracks are required to be 
identified as muons.

The cross-section for $J/\psi$
production is obtained from the ratio of peak to continuum production.
Assuming PDG values for $B_{\mu\mu}$ and ${\cal B}(J/\psi \to e^+ e^-)$, we obtain
$\Gamma(J/\psi \to e^+e^-) = 5.61 \pm 0.20 \kev$ and the full width of the $J/\psi$ to be
$\Gamma_{J/\psi} = 94.7 \pm 4.4 \kev$.

\subsection{Inclusive Analysis}

Alongside the exclusive analyses presented above, a fully inclusive analysis of hadronic 
ISR processes is being performed, with the goal of extracting $\Delta\alpha_{QED}^{Had}$ with
$3-4\%$ error. We select events with an ISR photon with center-of-mass energy greater than $3 \gev$.
 The various efficiency terms can all be
calibrated to $1\%$ or below; we have a triggering efficiency of $98\%$ and a fiducial photon 
detection efficiency of $90\%$. The $s'$ integrated luminosity spectrum can be computed from the BaBar 
integrated luminosity, which is known to about $1\%$. The precision of this calculation
is claimed to be less than $1\%$.

Leading background sources, such as radiative Bhabha, $e^+e^- \to \gamma\gamma$ and
virtual Compton scattering processes can be vetoed with minimal signal losses and biases. Other modes
such as $e^+e^- \to \mu^+\mu^- \gamma$ and $\tau^+\tau^-\gamma$  can be subtracted 
using theoretical predictions. Finally, $e^+e^- \to q\bar{q}$ events are a major source of background
for $s' > 5 \gev$, mainly due to production of high-momentum $\pi^0$ and $\eta$. Event shape variables of the
hadronic system can be used to suppress this background.

For this inclusive measurement, $s'$ is determined from the ISR photon energy.
Due to the EMC energy resolution of about $3\%$ for the energies considered here, the $R(s)/s$ spectrum
is distorted, especially at low $s'$. However, since $\Delta\alpha_{QED}^{Had}$ is expressed as an integral
in $R(s)/s$ with a weakly-varying weight factor, distortions in the spectrum do not affect the measured
value for $\Delta\alpha_{QED}^{Had}$. The energy resolution therefore
has minimal impact on the $\Delta\alpha_{QED}^{Had}$ measurement. The inclusive method cannot be
applied to the measurement for $a_{\mu}^{Had}$ since the weight factor in this case is strongly peaked
at $s' = 0$.

\section{Searches for Pentaquark Resonances}

Several experiments have recently claimed observations of exotic baryon resonances which
seem to be composed of 5 constituent quarks. The LEPS\cite{leps}
experiment has claimed observation of a resonance $\Theta^+$ at a mass of about $1540 \mev$.
The NA49 experiment\cite{na49} reports two degenerate states, $\Xi_5^0$ and $\Xi_5^{--}$ with masses
of $1862 \mev$. These resonances have been interpreted as members of a $\bar{10}+8$ multiplet of flavor $SU(3)$,
with the isospin-singlet $\Theta^+$ associated with states denoted as $N_5$, $\Sigma_5$ and $Xi_5$ 
in analogy with the usual baryon multiplets.

BaBar is well suited to search for these states, with excellent kaon and proton identification 
and excellent tracking resulting in good mass resolutions. Searches for the $\Theta^+$, $\Xi_5^0$, $\Xi_5^-$ 
and $\Xi_5^{--}$ states have been performed.

\begin{figure}
\begin{center}
\input{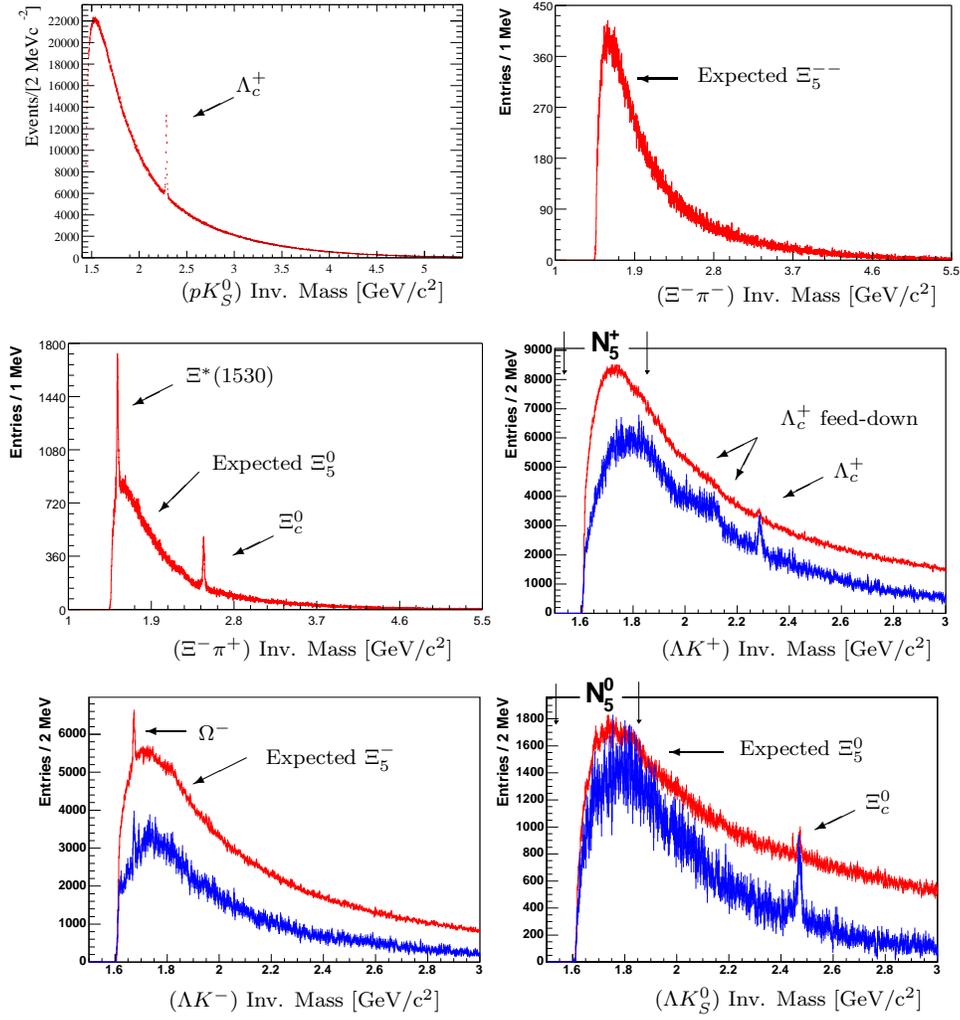}
\vspace{0.7cm}
\caption{\it Mass spectra for  $p K^0_S$ and $\Xi^{-}\pi^-$ (top row), $\Xi^+\pi^-$
and $\Lambda K^+$ (middle row) and  $\Lambda K^-$ and $\Lambda K^0_S$ (bottom row) 
using $123 \invfb$ of data. 
For the three latter plots, the upper and lower histograms correspond to $\Lambda K$
center-of-mass momenta respectively smaller than and greater than $3 \gev$, with the lower 
histogram scaled up by a factor of 10 for lisibility.
The positions of know resonances and expected $\Xi_5$ and $N_5$ pentaquarks are shown. 
}\label{spec}
\end{center}
%\vspace{-1.2cm}
\end{figure}

A search for $\Theta^+$  was done for the decay mode $\Theta^+ \to p K^0_S$. 
We expect a resolution of about $2 \mev$ on the $\Theta^+$ mass, which would be the most 
precise to date. However as shown in fig.~\ref{spec}, no peak is seen at the expected mass and only
a large signal for $\Lambda_C \to p K^0_S$ is observed.

A search for the $\Xi_5^0$ and $\Xi_5^{--}$ resonances was performed using the decay chain
$\Xi_5^{0/--}\to \Xi^- \pi^{\pm}$, $\Xi^- \to \Lambda \pi^-$, $\Lambda \to p\pi^-$, with the
proton identified as before.
As shown in fig.~\ref{spec}, no peak is seen at the expected masses. In the $\Xi^+\pi^-$ spectrum, prominent 
peaks for the $\Xi^*(1530)$ and $\Xi_c^0(2250)$ are seen. No structure is observed in the exotic
$\Xi^{-}\pi^-$ spectrum.

Searches for $\Xi_5^0 \to \Lambda K^0_S$, $N_5^0 \to \Lambda K^0_S$, $N_5^+ \to \Lambda K^+$
and $\Xi_5^- \to \Lambda K^-$, were also performed,
using kaon identification and reconstructing $\Lambda \to p\pi^-$ and $K^0_S \to \pi^+ \pi^-$
as above. As shown in fig.~\ref{spec} no exotic resonances were observed, while
sharp peaks for $\Omega^- \to \Lambda K^-$, $\Lambda_c^+ \to \Lambda K^+$ and $\Xi_c^0 \to \Lambda K^0_S$
are clearly seen.

\section{Conclusion}

Studies of hadronic cross-sections using initial-state radiation offer promising prospects
at BaBar. Many exclusive channels have already been measured, and more are in progress.
A fully inclusive analysis should also offer a precise measurement of $\Delta\alpha_{QED}^{Had}$.
Searches for pentaquark states have so far been negative, but they have served to highlight the potential
for the study of charmed and non-charmed baryons at high-luminosity $e^+e^-$ colliders.

\end{document}